\pdfoutput=1
\documentclass[11pt,twoside]{article}
\usepackage{asp2010}

\usepackage{longtable,lscape,graphicx,float,txfonts,natbib,multirow,color}

\def\HI{H\,{\sc i}}
\def\arcsec{$^{\prime}$$^{\prime}$}

\def\deg{$^{\circ}$}

\resetcounters

\newcounter{ppnum4}
\newcounter{ppnum5}
\begin{document}

\title{Molecular cloud determination in the Northern Galactic Plane}
\author{Joseph~C.~Mottram$^1$ and Chris.~M.~Brunt$^1$
\affil{$^1$School of Physics, University of Exeter, Stocker Road, Exeter, EX4 4QL, UK.}}

\begin{abstract}

The Exeter FCRAO CO Galactic Plane Survey consists of $^{12}$CO and $^{13}$CO (J=1$-$0) observations over the galactic plane covering 55\deg{}~$\leq$$\ell$$\leq$102\deg{},$\mid$$b$$\mid$~$\geq$~1\deg{} and 141\deg{}~$\leq$$\ell$$\leq$195\deg{}, -3.5\deg{}~$\leq$$b$$\leq$5.5\deg{} with a spatial resolution of $\sim$~45\arcsec{} and a spectral resolution of $\sim$~0.15kms$^{-1}$. We will present the methodology of a threshold-based cloud and clump determination method which retains hierarchical information, then discuss associating sources with clouds in the catalogue. Once complete, this catalogue of clouds and clumps will encompass the majority of the Northern Galactic Plane, providing knowledge of the molecular structure of the galaxy and the starting point for studies of the variation in star formation efficiency. In addition, it will allow us to identify clouds that have no or little star formation taking place inside them, which are often overlooked in the study of the conditions required for star formation to take place.

\end{abstract}

\section{Introduction}
\label{S:intro}

One of the principle objectives of any millimetre CO observations is to obtain the properties, such as size, velocity dispersion and mass, of the detected molecular structures \citep[e.g.][]{Brunt2003}. In data covering large regions in ($\ell$,$b$,$v$) space the initial identification of structures is non-trivial as there will be a range of scales on which emission is found and several independent regions may be detected along a given line-of-sight (LOS) due to the structure of the galaxy. Indeed substructure is also likely to be found on scales down to the spatial resolution of the observations which will probably not break into discrete populations. Another key objective is to be able to associate sources of emission at other wavelengths with CO emission structures so as to gain combined information about gas and continuum emission, and to pass distance and/or velocity information to 2-dimensional data \citep[e.g.][]{Urquhart2008b}. For ease of discussion we will use nomenclature similar to that of \citet{Williams2000} in that we define a cloud to be a region of molecular emission which is isolated at the lowest detectible level in the data, a clump to be an intermediate object in the heirarchical substructure within a cloud and a core to be a top level structure of a similar size to the beam.

The needs of the two objectives discussed above have often required different decisions regarding the identification of molecular structures. Obtaining cloud properties and/or understanding galactic scale structure (e.g. spiral arms) requires determination of low-level strucutures (i.e. clouds) but is less concerned with substructure. Source association requires good detection of dense structures (i.e. cores), but is often less concerned with intermediate and larger scale structures. Many previous studies have biased their structure determination towards the size scale they are interested in.

Given the continuous hierarchical nature of the structure within molecular clouds, any method for automatically identifying objects should take this into account in order that the results are neither biased towards or away from certain size scales or incomplete at a given scale. \citet{Rathborne2009} used a modified version of the CLUMPFIND algorithm \citep{Williams1994} with two different sets of input parameters in order to identify both clouds and clumps. However this did not truly track the hierarchy on all scales and was limited by the discrete way in which CLUMPFIND steps in temperature while searching for structures. \citet{Rosolowsky2008} developed a cloud decomposition method which follows the heirarchical structure of a region using dendrograms \citep[see e.g.][]{Houlahan1992} starting from each local maxima down to a common threshold until it merges with a neighbour, then considering the merged structures until a base threshold T$_{min}$ is reached. At each level objects which do not consist of a minimum number of pixels N$_{pix}$ within a given temperature difference $\Delta$T of the local peak temperature are rejected and, if connected at some lower threshold above T$_{min}$ to another local maxima are reassigned to that local maxima, similar to the criteria used by \citet{Brunt2003}.

\section{The Exeter-FCRAO CO Galactic Plane Survey}
\label{S:data}

\begin{figure}
\center
\includegraphics[width=1.0\textwidth]{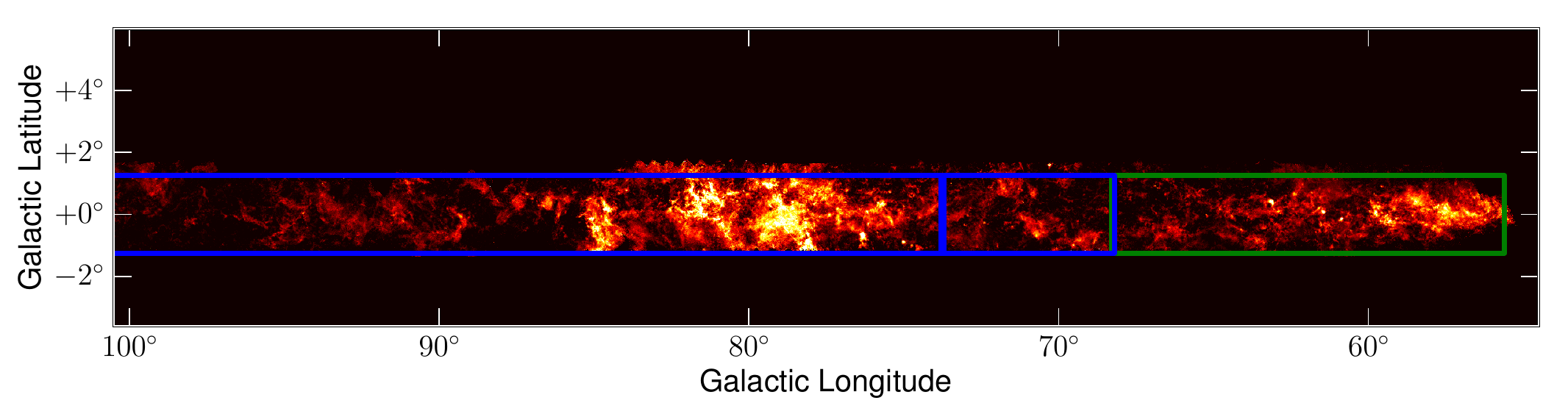}
\includegraphics[width=1.0\textwidth]{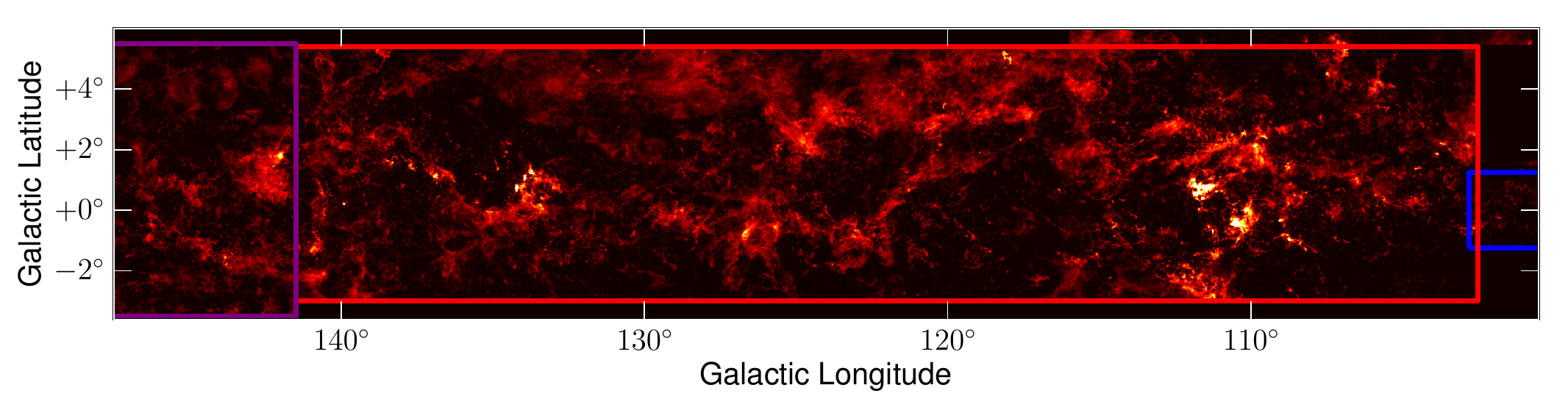}
\includegraphics[width=1.0\textwidth]{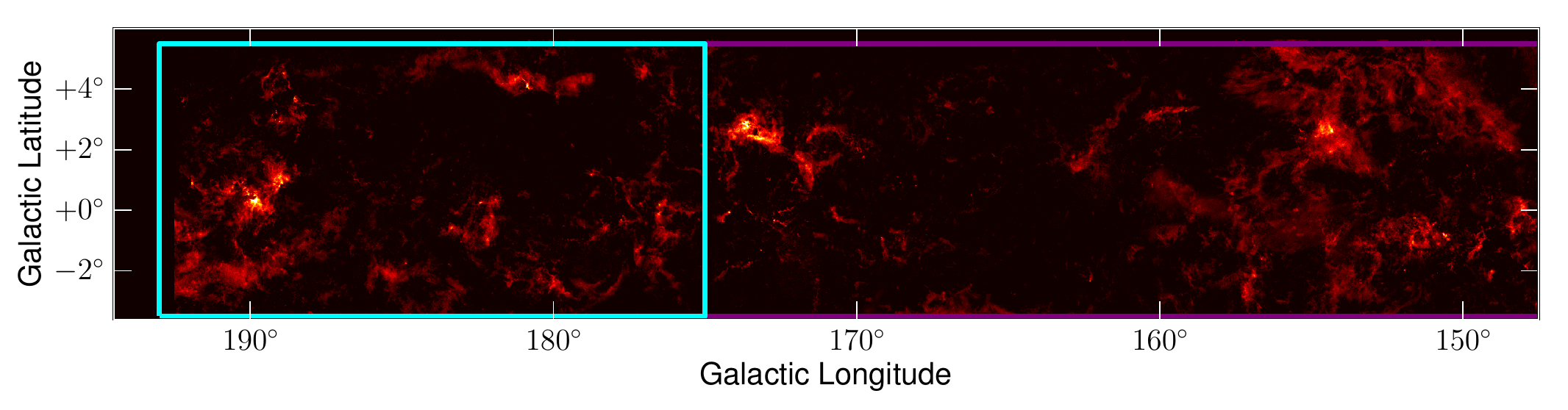}
\caption{Integrated maps of $^{12}$CO (J=1$-$0) emission as observed by the FCRAO between 55.6\deg{}$\leq\ell\leq$193\deg{} by the Exeter-FCRAO CO GPS and OGS. The boxes indicate the different regions covered: The Vulpecular Rift (green, 55.6\deg{}$\leq\ell\leq$68.3\deg{}, $\mid$$b$$\mid$$\approx$1.25\deg{}, -65kms$^{-1}$$\leq$$v$$\leq$65kms$^{-1}$), Cygnus (blue, 68.2\deg{}$\leq\ell\leq$73.8\deg{}, $\mid$$b$$\mid$$\approx$1.25\deg{}, -85kms$^{-1}$$\leq$$v$$\leq$45kms$^{-1}$ and 73.7\deg{}$\leq\ell\leq$102.8\deg{}, $\mid$$b$$\mid$$\approx$1.25\deg{}, -105kms$^{-1}$$\leq$$v$$\leq$25kms$^{-1}$), the OGS (red,102.5\deg{}$\leq\ell\leq$141.5\deg{}, -3.5\deg{}$\leq$$b$$\leq$5.4\deg{}, -150kms$^{-1}$$\leq$$v$$\leq$40kms$^{-1}$), the Extended Outer Galaxy (purple, 141.4\deg{}$\leq\ell\leq$175\deg{}, -3.5\deg{}$\leq$$b$$\leq$5.5\deg{}, -105kms$^{-1}$$\leq$$v$$\leq$25kms$^{-1}$) and the Anticentre (cyan, 175\deg{}$\leq\ell\leq$193\deg{}, -3.5\deg{}$\leq$$b$$\leq$5.5\deg{}, -65kms$^{-1}$$\leq$$v$$\leq$65kms$^{-1}$).}
\label{F:data_surveys}
\end{figure}

The Exeter-FCRAO CO Galactic Plane Survey (GPS) consists of $^{12}$CO and $^{13}$CO (J=1$-$0) fully sampled mapping observations obtained with SEQUOIA on the FCRAO with 45\arcsec{} and 46\arcsec{} spatial resolution respectively and $\sim$0.15kms$^{-1}$ spectral resolution over the regions 55.6\deg{}$\leq\ell\leq$102.5\deg{}, $\mid$$b$$\mid$$\approx$1.25\deg{} and 141.5\deg{}$\leq\ell\leq$193\deg{}, -3.5\deg{}$\leq$$b$$\leq$5.5\deg{}. This is broken up into roughly 4 regions (see Figure\ref{F:data_surveys}) with slightly different velocity ranges in order to follow galactic emission, and will be made public in the near future (Brunt, Heyer, Mottram $\&$ Douglas, in prep.). The region between 102.5\deg{}$\leq\ell\leq$141.5\deg{} is covered by the $^{12}$CO Outer Galaxy Survey \citep[OGS,][]{Heyer1998} which was observed with QUARRY on the FCRAO.

These data, though not covering as large an area, provide a factor of 10 improvement in spatial resolution and a factor of 4 in spectral resolution compared to those of \citet{Dame2001}. In addition, along with the Galactic Ring Survey \citep[GRS,][]{Jackson2006} they provide $^{13}$CO (J=1$-$0) observations for the vast majority of galactic plane in the first and second quadrants.

\section{A scheme for heirarchical object detection}
\label{S:decomp}

Our method for object detection has been developed from scratch, but has philosophical similarities to the approaches of \citet{Brunt2003} and \citet{Rosolowsky2008}. While the core methodology is general to any 3 dimensional intensity data, it is tailored to data in ($\ell$,$b$,$v$) space with intensities in temperature. 

We begin by thresholding the data at some minimum temperature T$_{min}$, usually defined by some integer value $m$ of the median $\sigma_{rms}$ noise of the data. Each contiguous region is checked to see that it passes a series of criteria, which are designed to ensure that only self-contained detections are accepted, rather than noise. These are that the region contains at least a minimum number of pixels N$_{pix}$ (usually calculated from the resolution of the data $\frac{4}{3}\pi\theta_{fwhm}^{2}\theta_{v}$), that the local maximum temperature T$_{max}$ within the region is detected above some difference from the threshold T$_{max}$-T$_{min}$~$>$~$\delta$T (usually an integer number $n$ times the median $\sigma_{rms}$ noise of the data) and that the intensity weighted centroid lies within the identified region. These regions are then termed the parent objects using tree terminology.

Following this, each region is examined to determine the minimum threshold between T$_{min}$ and the local T$_{max}$ where it divides into two or more separate objects, each of which satisfy the criteria discussed above. If this is the case, new objects are created which are daughters of the parent region under consideration, which is then removed from the list of objects to search and the search moves on to the next region. Alternatively, if the region does not divide to within $\delta$T of the local T$_{max}$, the region under consideration is the highest level detectable object relating to that local maximum. This process is repeated, including checking the daughter objects, until all regions are fully searched.

While this process is undertaken, a table is constantly updated which contains the cloud number and parent(s) and daughter(s) of every detected object. Next, for each parent cloud all top level daughters and the pixels assigned to lower objects in the local hierarchy within the parent are identified. Running from high to low temperature, each pixel is examined to see if any of the surrounding pixels are assigned to one of the top level daughters. If not, the next pixel is examined, if true then the pixel is reassigned to the top level daughter. This process is repeated until all pixels are assigned to a top level daughter. Where more than one top level daughter has a 'claim' on a pixel under consideration the daughter with the smallest temperature difference is chosen. This process, similar to the CLUMPFIND algorithm, is performed because the density structure of structures within molecular clouds is continuous, with dense cores 'owning' part of a lower density envelope which is shared with neighbouring cores. This bijection paradigm more accurately represents cloud and core properties than simple clipping them at a given threshold \citep[see][ for a more detailed discussion]{Rosolowsky2008}. 

This top level daughter deconvolution mask is then used to determine the properties of all cores, clumps and clouds. For clumps and cores at lower levels within the hierarchy, the pixels belonging to them are simply the sum of all pixels from the multiple top level daughters they contain.

\section{Applications}
\label{S:applications}

The following sections detail a few of the potential uses and applications of a large catalogue of molecular clouds, clumps and cores, aside from exploring the properties of the structures in their own right.

\subsection{Source Association}
\label{S:applications_association}

The association of sources detected in 2D imaging observations with molecular material enables velocity (and thus distance) information to be passed from the cloud to the sources associated with it. In addition, information about the star formation content and emission at other wavelengths of each molecular structure can be obtained. However, many of the current methods used to perform this association involve some degree of human decision making and hand-crafting \citep[e.g.][]{Chapin2008}, due to the fact that there can often be more than one molecular structure along a given LOS.

Given the scale of both the region observed by the Exeter-FCRAO CO GPS and current infrared, sub-mm and mm surveys \citep[e.g.][]{Churchwell2009,Moore2005,Molinari2010} a more automated approach is required. \citet{Brunt2003} calculated the expectation N$_{E}$ that a given location will be within $\delta$r of emission with peak temperature T$_{P}$ by seeing the fraction of randomly located test particles that fulfil this criteria. \citet{Kerton2003} then used this expectation to find the most likely association between the clouds identified by \citet{Brunt2003} and IRAS point source catalogue \citep[PSC,][]{Beichman1988} sources.

\begin{figure}
\center
\includegraphics[width=0.49\textwidth]{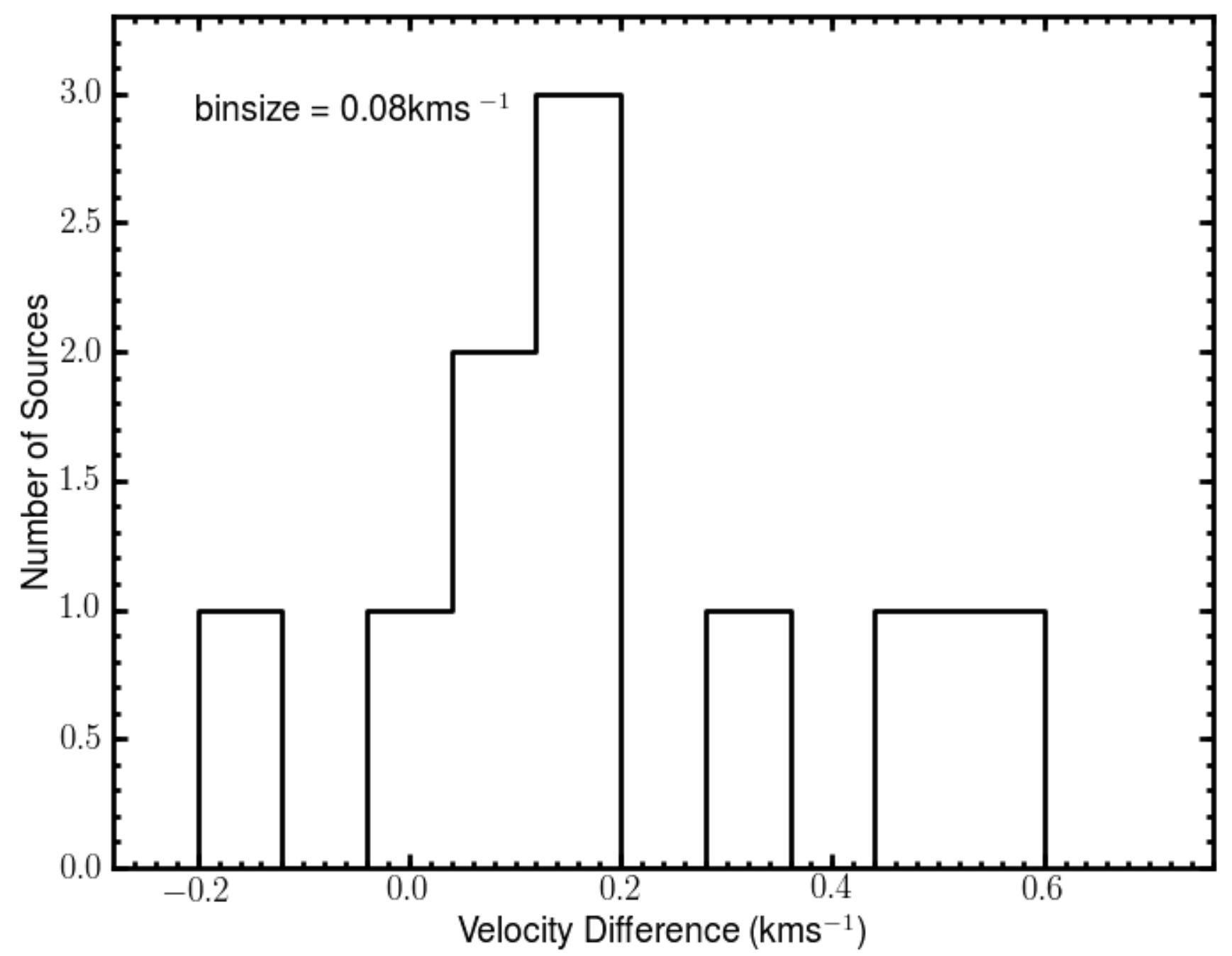}
\includegraphics[width=0.49\textwidth]{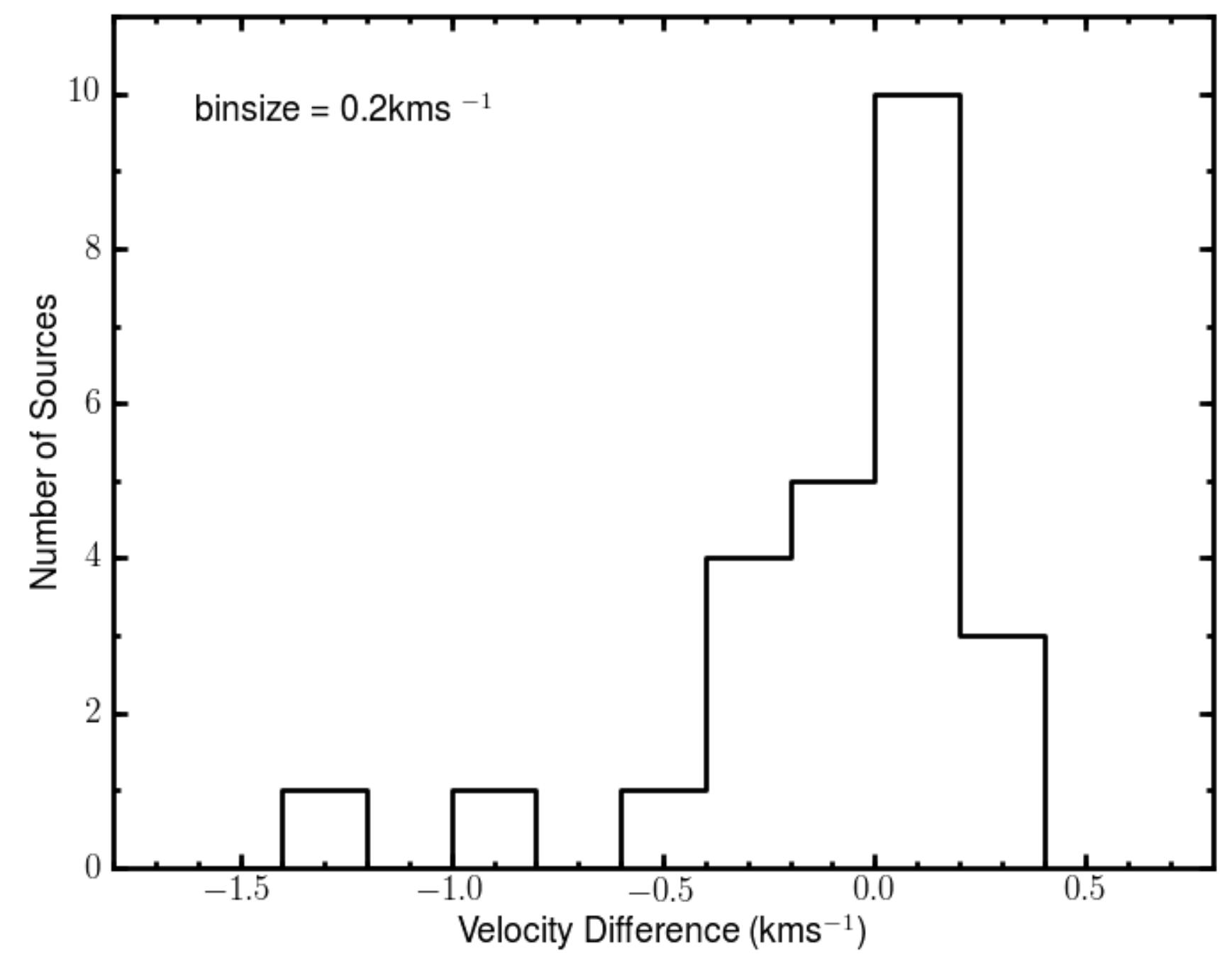}
\caption{The velocity difference between velocities associated with BLAST \citep{Chapin2008} and RMS \citep{Urquhart2008b} source with those assigned by associating the sources to molecular structures in a heirarchical cloud catalogue using the methodology of \citet{Kerton2003}.}
\label{F:applications_association_comparison}
\end{figure}

Adopting this approach, we compare results obtained using this method with those of two more hand-worked studies in a region around $\ell$=59\deg{}. \citet{Chapin2008} used an early reduction of our CO data to obtain velocities for their Balloon-borne Large Aperture Submillimeter Telescope \citep[BLAST,][]{Pascale2008} sub-mm point sources. The Red MSX Source (RMS) Survey, which is a multi-wavelength search for young massive stars throughout the galactic plane \citet{Hoare2005,Urquhart2008a,Mottram2010}, performed their own pointed $^{13}$CO observations to their sources in this region \citep{Urquhart2008b} in order to obtain kinematic information for their sources. The difference between our results using the method discussed above and those obtained by \citet{Chapin2008} and \citet{Urquhart2008b} are shown in figure~\ref{F:applications_association_comparison}. These are small and occur mainly due to the fact that we assign the velocity obtained from the centroid of the cloud to the source, while the original determinations used the CO velocity along the line of sight of the source. The latter approach has been know to cause large differences in derived velocity between nearby sources which are known to reside in the same complex. For example, \citep{Chapin2008} find a velocities in the range 26$-$31kms$^{-1}$ for sources associated with a molecular complex in the Saggitarius spiral arm for which they set a common distance. The method of \citet{Kerton2003} automatically assigns the same velocity to sources associated with the same molecular structure.

The ability to associate sources with clouds will allow us to obtain the star formation efficiency of the clouds and explore how this varies across the galactic plane. It will also allow us to identify those clouds which, despite having sufficient density, have little or no star formation taking place within them. The relative frequency of such clouds will provide important clues as to how quickly star formation begins with molecular clouds, and provide important targets for studies of the earliest stages of star formation.

\subsection{Distance Assignment}
\label{S:applications_distances}

While kinematic distances can be obtained to all objects identified in the molecular data, the uncertainties associated with this approach are often quite large. An alternative approach which is possible with a heirarchical catalogue of molecular structures is to associate sources with distances obtained by more accurate methods \citep[e.g. maser parallax,][]{Reid2009} to clouds, clumps or cores within the molecular material, then allow this distance to be passed on to all connected parts of the hierarchy above the local minimum confusion threshold (LMCT). The LMCT is simply the level in the hierarchy at which more than one conflicting distance measurement is available. As improved distance information becomes available, the distances passed to the molecular structures and the other sources associated with then can be updated as well, without having to revisit the CO decomposition itself. This will also ensure that complexes have the same distance, which may not be the case if kinematic distances are used, as discussed above.

\subsection{Comparison with \HI{}}
\label{S:applications_HI}

From the regions identified in the deconvolution masks, other ($\ell$,$b$,$v$) data sets, such as \HI{} data from the Canadian GPS \citep[CGPS,][]{Taylor2003} can be compared to the clouds, clumps and cores identified in the CO data. This would enable the kinematic distance ambiguity to be resolved for these structures in the inner galaxy using \HI{} self absorption techniques. It will also allow cross-comparison with catalogues of \HI{} absorption and emission features \citep{Gibson2005a,Gibson2005b}, so that large statistical samples of clouds can be selected to help study the interplay between neutral and molecular material in our galaxy.

\section{Summary}
\label{S:summary}

We have a method for identifying heirarchical structure within molecular $\ell$-$b$-$v$ observations, which will be applied to the Exeter-FCRAO CO Galactic Plane Survey. The catalogue produced by this process will be associated with sources identified at other wavelengths, which will allow distances from various methods to be passed to the clouds, clumps and cores within it. What is more the star formation efficiency of the molecular structures will be obtained and variations explored. Other applications beyond exploring the directly measured properties of clouds within the catalogue include comparison with \HI{} emission and absorption structures, comparison of distance ambiguity resolution using extinction and \HI{} self absorption and obtaining the Schmidt-Kennicutt law within the galactic plane.

\acknowledgements This work was supported by the Science and Technologies Research Council of the United Kingdom (STFC) Grant ST/F003277/1 to the University of Exeter, Marie Curie Re-Integration Grant MIRG-46555, and NSF grant AST 0838222 to the Five College Radio Astronomy Observatory. J.C.M. is supported by a Postdoctoral Research Associate grant from STFC. C.B. is supported by an RCUK fellowship at the University of Exeter, UK. The Five College Radio Astronomy Observatory was supported by NSF grant AST 0838222. This paper made use of information from the Red MSX Source survey database at www.ast.leeds.ac.uk/RMS which was constructed with support from the Science and Technology Facilities Council of the UK.

\end{document}